\documentclass[prd,aps,nofootinbib,floats,letterpaper,floatfix,superscriptaddress,groupedaddress,twocolumn,showpacs]{revtex4-1}

\usepackage[dvips]{graphicx,color}
\usepackage{hyperref}

\begin{document}

\title{The capture of dark matter particles through the evolution of low-mass stars}

\author{Il\'idio Lopes}
\email[]{ilidio.lopes@ist.utl.pt}

\affiliation{Departamento de F\'isica, Universidade de \'Evora, Portugal}
\affiliation{CENTRA, Instituto Superior T\'ecnico, Lisboa, Portugal}

\author{Jordi Casanellas}
\email[]{jordicasanellas@ist.utl.pt}

\author{Daniel Eug\'enio}
\email[]{dan.far.eugenio@gmail.com}

\affiliation{CENTRA, Instituto Superior T\'ecnico, Lisboa, Portugal}

\date{\today}

\begin{abstract}
We studied the rate at which stars capture dark matter (DM) particles, considering different assumptions regarding the DM characteristics and, in particular, investigating how the stellar physics influences the capture rate. Two scenarios were considered: first, we assumed the maximal values for the spin-dependent and spin-independent DM particle-nucleon scattering cross sections allowed by the limits from direct detection experiments. Second, we considered that both scattering cross sections are of the same order, with the aim of studying the dependencies of the capture rate on stellar elements other than hydrogen. We found that the characteristics of the capture rate are very different in the two scenarios. Furthermore, we quantified the uncertainties on the computed capture rate ($C_{\chi}$) and on the ratio between the luminosities from DM annihilations and thermonuclear reactions ($L_{\chi}/L_{nuc}$) derived from an imprecise knowledge of the stellar structure and DM parameters. For instance, while an uncertainty of 10\% on the typical DM velocity leads to similar errors on the computed $C_{\chi}$ and $L_{\chi}/L_{nuc}$, the same uncertainty on the stellar mass becomes more relevant and duplicates the errors. Our results may be used to evaluate the reliability of the computed capture rate for the hypothetical use of stars other than the Sun as DM probes. 
\end{abstract}


\maketitle

\section{Introduction\label{sec-intro}}
The study of the rate at which stars capture dark matter (DM) particles is of vital importance to understand in which situations stars are able to accumulate enough DM to influence their evolution. The possibility of using the properties of stars within dense DM halos as an indirect method to investigate the nature of DM relies on the precision of the capture rate calculation. This quantity depends on both the DM characteristics and the details of the stellar structure \citep{art-PressSpergel1985,art-GriestSeckel1987NuPhB}. 

In the case of the Sun, a precise calculation of the capture rate is very important to predict the neutrino flux from DM annihilations in the center of the star \citep{art-WikstromEdsjo2010JCAP,art-Ellisetal2010PhRevD,art-EsmailiFarzan2010PhRvD,art-DemidovSuvorova2010JCAP} and to calculate the changes in the solar neutrino flux induced by an isothermal core created by the energy transport due to DM particles conduction \citep{let-LopesSilk2002,art-FrandsenSarkar2010PhRvL,art-Taosoetal2010PhRvD,art-LopesSilk2010Science}. In this context, the systematical errors in the determination of the local DM density were recently studied \citep{art-CatenaUllio2010JCAP,art-WeberdeBoer2010AA,art-Patoetal2010PhRevD}, as well as the uncertainties coming from other astrophysical sources, as the shape of the velocity distribution of the DM particles or the motion of the Sun in respect to the DM halo \citep{art-SerpicoBertone2010PhRevD,art-Kuhlenetal2010JCAP}. These works have shown that the systematic errors introduced by such astrophysical parameters are considerably large if one wants to extract information about the type of DM particle only from current direct or indirect detection experiments.

On the other hand, the scope of our work is to characterize the capture rate for stars other than the Sun. Recent works have shown that, when embedded in dense halos of DM, stars may dramatically change their properties \citep{let-Spolyaretal2008,art-Iocco2008,art-BertoneFairbairn2008,art-FreeseGS2008,art-Ioccoetal2008,art-SchleicherBK2008,art-HooperSpolyaretal2010PhRevD,art-RipamontiIocooetal2010MNRAS,art-GondoloHuhKimScopel2010JCAP,art-SivertssonGondolo2010ApJ,art-ZackScottIoccoetal2010ApJ}. In these cases, the uncertainties in the knowledge of the typical parameters governing the capture rate are much larger. Generally, in the literature, when the capture rate of DM particles  is calculated for stars other than the Sun, as for compact stars \citep{art-Kouvaris2008,art-KouvarisTinyakov2010PRD,art-LavallazFairbairn2010PhRvD,art-MoskalenkoWai2007} or low-mass stars \citep{art-Scottetal2007,art-Fairbairnetal2008,art-CasanellasLopes2009ApJ}, the fiducial values for the local Keplerian velocity ($v_{\star}=220\;$km s$^{-1}$) and DM velocity distribution (Maxwell-Boltzmann [MB] distribution with a velocity dispersion $\bar{v_{\chi}}=270\;$km s$^{-1}$) are assumed. However, in the situations where these stars can exist, these parameters may have very different values. For instance, in a possible interesting place such as near the center of our Galaxy, the velocities of the stars range from 10 to 500 km s$^{-1}$ \citep{art-LuGhezetal2009} and the DM particles may have motions dominated by the gravitational potential of the hypothetical central black hole \citep{chap-Merrit2010}. Simultaneously, the stellar velocity dispersions measured in nearby galaxies range from 10 to 400 km s$^{-1}$ \citep{art-Hoetal2009ApJS}. In the first part of this paper we explore how the stellar capture rate changes with the astrophysical parameters and DM characteristics in order to grasp the possible modifications in the effects that DM annihilation may have on stars other than the Sun.   

In the second part of this paper we characterize how the capture rate changes during the life of a star (from the collapse of the protostar to the helium flash) considering stars with different masses (0.5 M$_{\odot}$ to 7 M$_{\odot}$) and metallicities (Z=0.0004 to Z=0.04).

We will consider two scenarios. First, a scenario where the capture is dominated by the spin-dependent (SD) collisions of hydrogen atoms with the DM particles, which corresponds to assuming the maximal DM particle-nucleon scattering cross sections allowed by the limits from direct detection experiments. Second, a scenario where the SD and spin-independent (SI) scattering cross sections are of the same order, a plausible possibility given that both interactions came from similar processes \citep{art-GoodmanWitten1985PhysRevD,art-Griest1988PhysRevD}. In fact, in many supersymmetric models the scalar interaction (SI) often dominates the elastic scattering \citep{art-Bednyakovetal1994PhRevD,rev-JungmanKamionkowskiGriest1996}. Within this assumption, other stellar elements such as oxygen, helium, or iron arise as the more relevant ones in capturing DM particles. Thus, we also explore how different stellar and DM physics change the role of the dominant elements in the capture rate. Finally, in the last part of this paper we study how the uncertainties in the determination of these parameters influence the computed capture rate and the impact of the annihilation of DM particles inside stars.

\section{Stellar capture of DM particles}
To study the various dependencies of the capture rate some routines of the {\sf DarkSUSY} code \citep{art-GondoloEdsjoDarkSusy2004} were adapted in order to include them on a modified version of the stellar evolution code {\sf CESAM} \citep{art-Morel1997}. The latter code has a very refined stellar physics, tested against helioseismic data in the case of the Sun \citep{art-CouvidatTuChi2003,art-Turck-Chiezeetal2010ApJ}. If not stated otherwise, we assume a stellar metallicity $Z=0.019$, an helium mass fraction $Y=0.273$, and abundances of the other elements as the solar ones \citep{art-AsplundGrevesseSauval2005}.

The capture rate is computed in our code according to the expressions of Gould \citep{art-Gould1987},
\begin{equation}
    C_{\chi}(t) = \sum_i \int^{R_\star}_0 4\pi r^2\int^\infty_0 \frac{f_{v_{\star}}(u)}{u}w\Omega_{v,i}^-(w)\,\mathrm{d}u\,\mathrm{d}r\;,
\label{eq-cap}
\end{equation}
\begin{equation}
 \Omega_{v,i}^-(w) = \frac{\sigma_{\chi,i} n_i(r)}{w}\Big(v_{e}^2-\frac{\mu_{-,i}^2}{\mu_i}u^2\Big)\theta\Big(v_{e}^2-\frac{\mu_{-,i}^2}{\mu_i}u^2\Big),
\end{equation}
\begin{equation}
    \mu_i\equiv\frac{m_\chi}{m_{\mathrm{n},i}}, \quad\mu_{\pm,i}\equiv\frac{\mu_i\pm 1}{2}\;,
\end{equation}
where

$\Omega_{v,i}^-(w)$ is the rate of scattering of a DM particle with the nucleus of an element $i$, from an initial velocity $w$ at the radius of the collision to a velocity lower than the escape velocity of the star $v_{e}(r)$ at that radius (kinetic factor);

$f_{v_{\star}}(u)$ is the velocity distribution of the DM particles seen by the star, which depends on the velocity of the star $v_{\star}$ and on the velocity distribution of the DM particles in the halo $f_0(u)$;

$m_{\chi}$ is the mass of the DM particle;

$\sigma_{\chi,i}$ is its scattering cross section with an element $i$, which is: $\sigma_{\chi,i}=\sigma_{\chi,SI} A_i^2 \Big(\frac{m_\chi m_{\mathrm{n},i}}{m_\chi+m_{\mathrm{n},i}}\Big)^2 \Big(\frac{m_\chi+m_{p}}{m_\chi m_{p}}\Big)^2$ for all stellar elements except for hydrogen, which has also the contribution from the SD interactions $\sigma_{\chi,H}=\sigma_{\chi,SI}+\sigma_{\chi,SD}$;

$m_{\mathrm{n},i}$, $A_i$ are the nuclear mass and the atomic number of the element $i$;

$n_i(r)$ is the density of the element $i$ at a radius $r$; and

$R_\star$ is the total radius of the star.

For stellar elements other than hydrogen a suppression form factor is considered, along the lines of Gould \citep{art-Gould1987}, to account for the influence of the size of the nucleus on the interactions. Thus, the scattering rate is
\begin{small}
\begin{eqnarray}
\Omega_{v,i}^-(w) = \frac{\sigma_{\chi,i} n_i(r)}{w} \frac{2 E_0}{m_{\chi}} \frac{\mu_{+,i}^2}{\mu_i} \left\{\exp\left(-\frac{m_{\chi} u^2}{2 E_0}\right)\right.-\nonumber\\
\exp\left(-\frac{m_{\chi} u^2}{2 E_0}\frac{\mu_i}{\mu_{+,i}^2}\right) 
\left.\exp\left(-\frac{m_{\chi} v_{e}^2}{2 E_0} \frac{\mu_i}{\mu_{-,i}^2} \left(1-\frac{\mu_i}{\mu_{+,i}^2}\right)\right) \right\},
\end{eqnarray}\end{small}
where $E_0\simeq 3\hbar / (2m_{\mathrm{n},i}(0.91m_{\mathrm{n},i}^{1/3}+0.3)^2$ is the characteristic coherence energy. The abundances of $^{2}$H, $^{4}$He, $^{12}$C, $^{14}$N, $^{16}$O and other isotopes which are produced or burned during the proton-proton chain, CNO cycle, or triple alpha nuclear reactions are followed by our code. For iron, neon, and silicon, among others, their proportion over the remaining mass is set as in the solar composition.

The new energy transport mechanism by conduction of the DM particles \citep{art-FaulknerGilliland1985ApJ} and the new energy source by the annihilation of DM particles inside the star \citep{art-SalatiSilk1989} are also included in this version of the code. However, these processes do not influence the total capture rate of the stars computed in this work.

\section{Capture rate dependence\\on DM properties
\label{sec-exploreDM}}
\subsection{DM halo density and scattering cross sections}
\begin{figure}[!t]
\centering
 \includegraphics[scale=0.95]{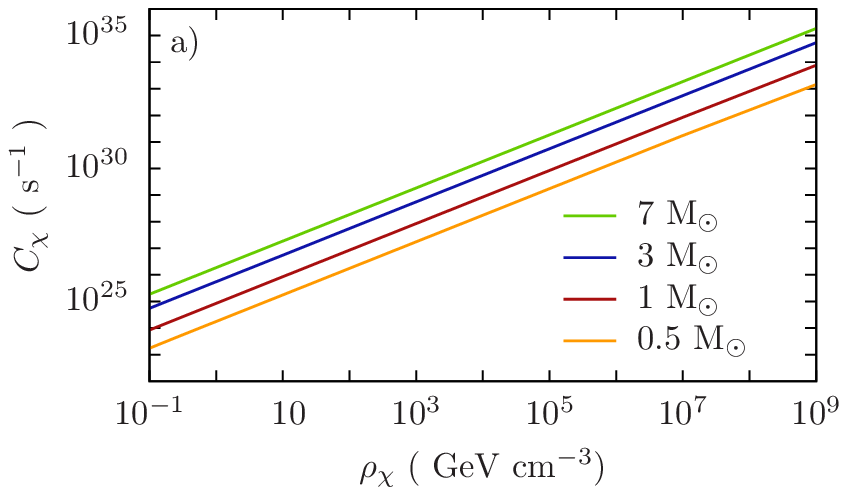}\\
 \includegraphics[scale=0.95]{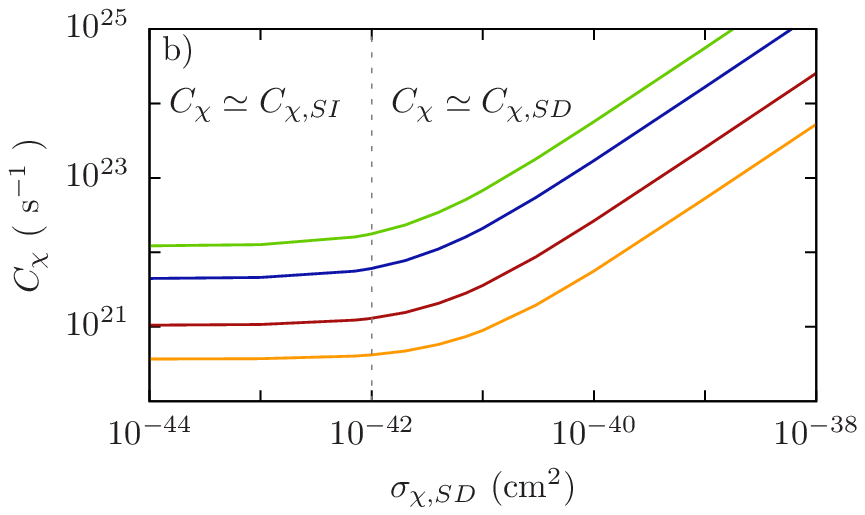}\\
 \includegraphics[scale=0.95]{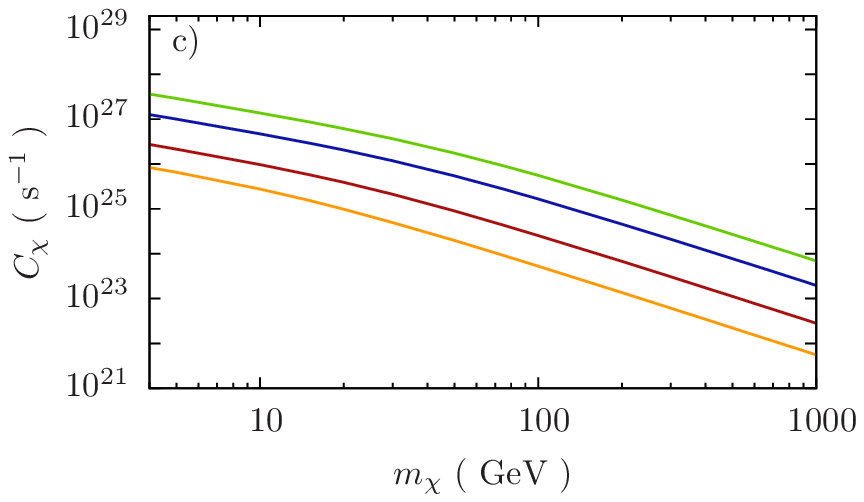}
 \caption{Rate at which DM particles are captured for stars of different masses, considering different DM halo densities (a), different SD DM particle-nucleon scattering cross sections (b), and different masses of the DM particles (c). If not stated otherwise, a halo of DM particles with $\rho_{\chi}=0.3\;$GeV$\;$cm$^{-3}$, $m_{\chi}=100\;$GeV and scattering cross sections $\sigma_{\chi,SD}=10^{-38}\;$cm$^2$ and $\sigma_{\chi,SI}=10^{-44}\;$cm$^2$ is  assumed.}
\label{fig-dens-sigSD-mx}
\end{figure}
The total number of DM particles captured by a star is proportional to both the density of DM in the halo $\rho_\chi$ and the DM particle-nucleon scattering cross section $\sigma_{\chi}$ [see Fig. \ref{fig-dens-sigSD-mx}(a) and 1(b)]. Hence, all the capture rates that will be shown in this work may be simply rescaled if the reader wants to consider other values of $\rho_\chi$ or $\sigma_{\chi}$. If not stated otherwise, a DM density $\rho_{\chi}=0.3\;$GeV$\;$cm$^{-3}$ and DM-nucleon scattering cross sections $\sigma_{\chi,SD}=10^{-38}\;$cm$^2$ \citep{art-PICASSOSD2009,art-COUPP2011PhRvL} and $\sigma_{\chi,SI}=10^{-44}\;$cm$^2$ \citep{art-XENON10_SI2008} (the largest cross sections allowed by the limits from direct detection experiments) are assumed in our computations, as is generally done in the literature when the effects of DM particles on stars are studied \citep{art-Taosoetal2008PhRevD,art-YoonIoccoAkiyama2008}. Within this assumption, the capture rate is always dominated by the contribution of the SD collisions of the DM particles with hydrogen atoms.

On the other hand, the dependencies of the capture rate change when values for the SD scattering cross section closer to the SI ones are considered. We found that for $\sigma_{\chi,SD}$ smaller than $10^{-42}\;$cm$^2$ the SI interactions are responsible for most of the captures [see Fig. \ref{fig-dens-sigSD-mx}(b)]. More generally, for a SD scattering cross section smaller than a hundred times the SI one, the SI collisions dominate the total capture rate. In this scenario, other stellar elements, such as oxygen, iron, or helium, play an important role in the capture of DM particles. This situation is studied in-depth in Sec. \ref{sec-stellphys}.

We note that, for stellar metallicities different from the solar one, the ratio $r\equiv\sigma_{\chi,SD}/\sigma_{\chi,SI}$ below which SI interactions dominate changes: $r\simeq70$ for Z=0.0004 while $r\simeq1000$ for Z=0.04.

\subsection{Mass of the DM particles}
\begin{figure}[!t]
\centering
 \includegraphics[scale=0.95]{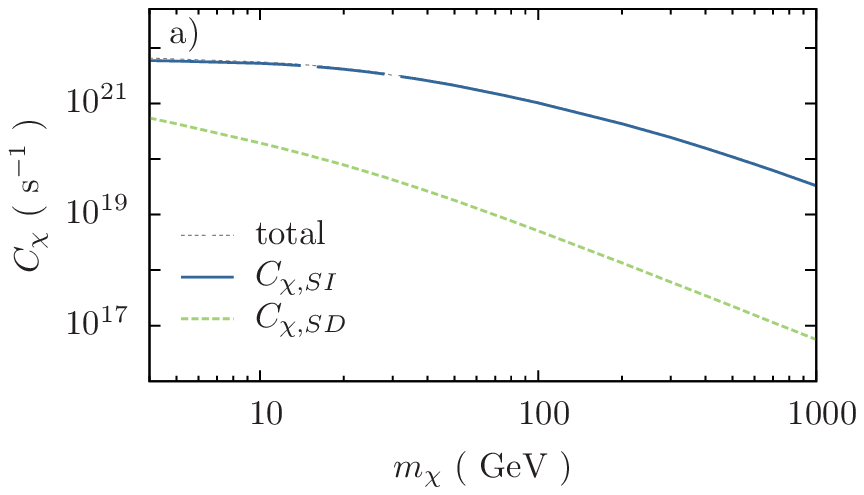}
 \includegraphics[scale=0.95]{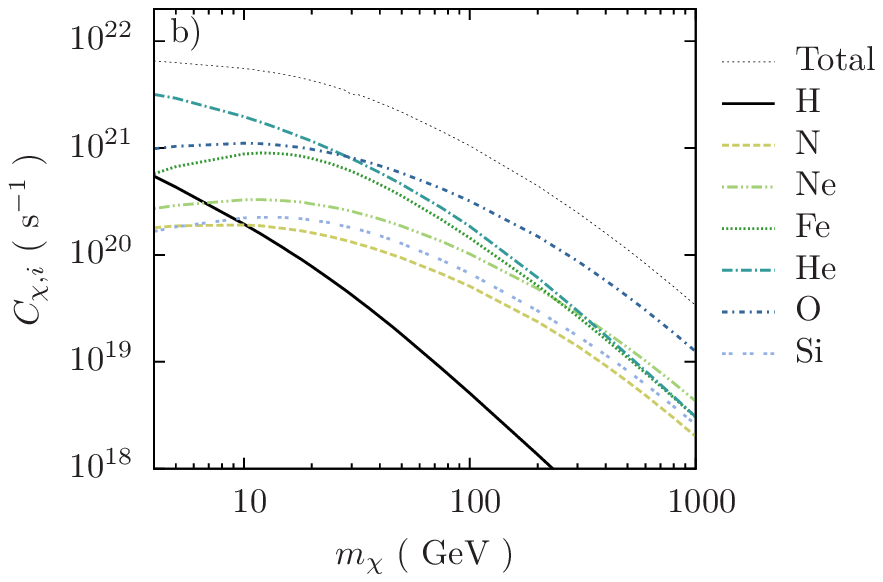}
 \caption{(a) Rate at which DM particles are captured for a 1 M$_{\odot}$ MS star due to the SD interactions of the DM particles with hydrogen (green dashed line) and due to SI interactions with hydrogen, nitrogen, neon, iron, helium, oxygen, and silicon, among others (blue continuous line). (b) Capture rate discriminated by the element responsible for the collision that led to the capture. We assumed a halo of DM particles with $\rho_{\chi}=0.3\;$GeV$\;$cm$^{-3}$ and the DM-nucleon scattering dominated by the spin-independent (SI) component, $\sigma_{\chi,SI}=\sigma_{\chi,SD}=10^{-44}\;$cm$^2$.}
\label{fig-mxSISD}
\end{figure}
The capture rate is roughly inversely proportional to the mass of the DM particles $m_{\chi}$, as it is proportional to the number density of DM particles in the halo $\frac{\rho_\chi}{m_\chi}$. In Fig. \ref{fig-dens-sigSD-mx}(c) are shown the big decreases found in $C_{\chi}$ when $m_\chi$ goes from 4 to 1000 GeV. We have chosen a range of masses above the limit from which evaporation can be considered negligible \citep{art-SpergelPress1985,art-GriestSeckel1987NuPhB,art-Gould1987ApJI}, which includes the light weakly interacting massive particles (WIMPs) recently invoked as the DM candidates that can reconcile the results from different direct detection experiments \citep{art-Savageetal2009JCAP,art-Fitzpatricketal2010PhRvD}. 

The drop in the capture rate due to a large $m_\chi$ has no consequences when considering the effects of DM annihilation inside stars. When $m_\chi$ is large, the star captures a small number of DM particles, but each of the few annihilations that take place release more energy, compensating for the low capture. On the other hand, considering a different $m_\chi$ does influence the distribution of DM particles inside the star. This fact has consequences on the seismological signature of the isothermal core created in the center of the Sun by the transport of energy through DM conduction \citep{art-LopesSH2002,art-LopesBS2002,art-Bottinoetal2002,art-Cumberbatchetal2010PhRevD,art-LopesSilk2010ApJL}, and on the strong seismological signature of DM annihilation inside solarlike stars within very dense DM halos \citep{art-CasanellasLopes2010MNRAS}.

Alternatively, in the scenario where the capture rate is dominated by the SI interactions the drop on the capture rate when the DM mass increases is not so steep [see Fig. \ref{fig-mxSISD}(a)]. This is a consequence of the capture due to the collisions of the DM particles with the heavier elements. These interact through SI scattering, while hydrogen, the lightest element, is the only one contributing to the SD capture. The capture rate of DM particles with different masses discriminated by the elements that are responsible for the collisions that lead to the capture, $C_{\chi,i}$,  is shown in Fig. \ref{fig-mxSISD}(b). While $^4$He dominates the capture of lighter WIMPs, $^{16}$O does the same for the heavier ones \citep{Edsjo-workinprogress}. In Fig. \ref{fig-mxSISD}(b) it can also be seen that each of the elements has a peak of its capture rate when the WIMP mass is roughly equal to its own mass \citep{art-GriestSeckel1987NuPhB,art-Gould1987}. Therefore, while the captures due to the hydrogen and helium are highly suppressed for larger DM masses, the capture for heavier elements decreases less steeply with $m_\chi$.

The causes for the enhancement or suppression of the $C_{\chi,i}$ at different DM masses are found in three different factors, all of them functions of $m_\chi$: the SI scattering cross section, the kinetic factor and the form factor. Both of the first two factors introduce an $A_i^2$ dependence on $C_{\chi,i}$, thus enhancing the capture rate due to collisions with the heavier elements. On the other hand, the nuclear form factor slows down this effect suppressing the capture rate only for the isotopes with larger atomic numbers (see Ref. \cite{art-Gould1987}).

\subsection{Phase space of the DM particles}
\begin{figure}[!t]
\centering
 \includegraphics[scale=0.95]{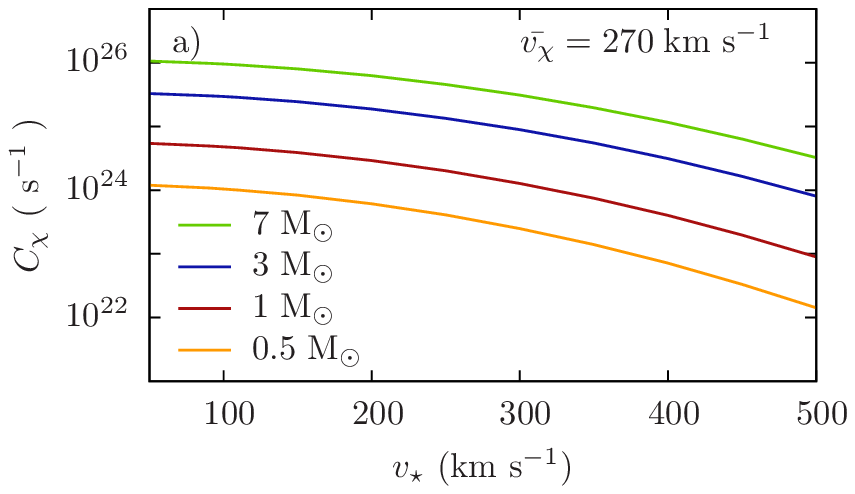}
 \includegraphics[scale=0.95]{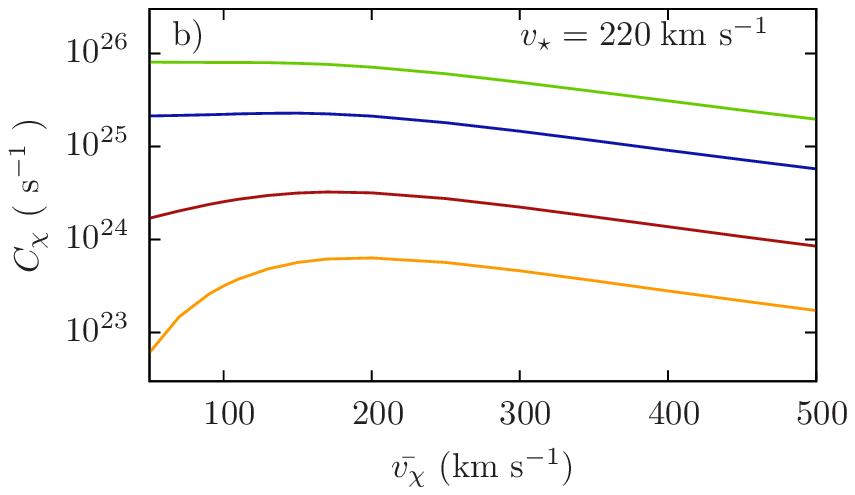}
 \includegraphics[scale=0.95]{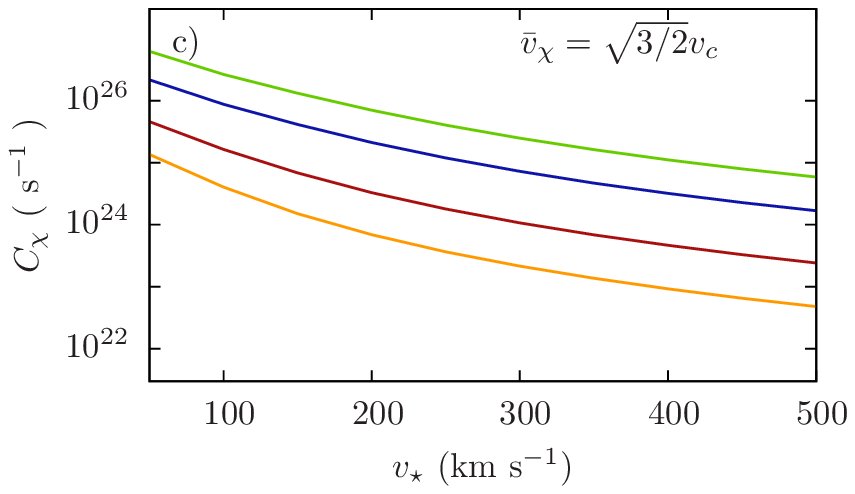}
 \caption{Rate at which DM particles are captured for stars of different masses, considering different stellar velocities (a), different DM typical velocities (b), and varying both speeds relating them through $\bar{v}_{\chi}=\sqrt{3/2}v_c$ (c). We assumed a halo of DM particles with $\rho_{\chi}=0.3\;$GeV$\;$cm$^{-3}$, $m_{\chi}=100\;$GeV and the DM-nucleon scattering dominated by the spin-dependent (SD) component, $\sigma_{\chi,SD}=10^{-38}\;$cm$^2$.}
\label{fig-velocities}
\end{figure}

Generally, the literature assumes a Maxwell-Boltzmann distribution for the velocities of the DM particles $f_0(u)$, with a dispersion $\bar{v_{\chi}}$, leading to a velocity distribution seen by the star of \citep{art-Gould1987,art-Scottetal2009MNRAS,art-SivertssonEdsjo2010PhRvD}:\begin{equation}
    f_{v_{\star}}(u) = f_0(u) \exp\Big(-\frac{3v_\star^2}{2\bar{v_{\chi}}^2}\Big) \frac{\sinh(3uv_\star/\bar{v_{\chi}}^2)}{3uv_\star/\bar{v_{\chi}}^2}
\label{eq-fvs}
\end{equation}
\begin{equation}
    f_0(u) = \frac{\rho_\chi}{m_\chi} \frac{4}{\sqrt{\pi}} \Big(\frac{3}{2}\Big)^{3/2} \frac{u^2}{\bar{v_{\chi}}^3} \exp\Big(-\frac{3u^2}{2\bar{v_{\chi}}^2}\Big).
\label{eq-fv0}
\end{equation}
Within this assumption, we explored how the capture rate changes for different values of $v_{\star}$ and $\bar{v_{\chi}}$. First, a MB distribution of the DM particles with a fixed $\bar{v_{\chi}}=270\;$km s$^{-1}$ was considered and the stellar velocity $v_{\star}$ was varied from 50 to 500 km s$^{-1}$ [see Fig. \ref{fig-velocities}(a)]. We found that at high stellar velocities the capture rate drops because the DM particles that the star encounters are more energetic and consequently are more difficult to capture. Second, a fiducial value for $v_{\star}=220\;$km s$^{-1}$ was considered and the dispersion velocity of the DM particles $\bar{v_{\chi}}$ was varied from 50 to 500 km s$^{-1}$. As expected, for higher dispersions of the DM velocity distribution the capture rate is lower, as more DM particles have high velocities and are not captured. We note that in this situation one may consider to truncate the velocity distribution at the galactic escape velocity. This was included in the capture rate computed by Ref. \cite{art-Scottetal2009MNRAS} in the case of main sequence (MS) stars at the Galactic center and by Ref. \cite{art-SerpicoBertone2010PhRevD} in the case of the Sun. The latter authors found that the uncertainties in the knowledge of the local escape velocity lead to errors on the estimation of the solar capture rate of approximately 10\%.

Assuming an isotropic, Gaussian velocity distribution of the DM particles, the velocity dispersion can be related to the circular speed (the velocity that a mass would have on a circular orbit in the galactic plane) using the Jeans equation \citep{book-BinneyTremaine1987}, leading to $\bar{v}_{\chi}=\sqrt{3/2}v_c$. We considered the case of stars with $v_{\star}=v_c$ (an assumption that in the case of the Sun introduces an error of $\sim10\%$ \citep{art-SerpicoBertone2010PhRevD}) within DM halos with velocity dispersions $\bar{v}_{\chi}=\sqrt{3/2}v_{\star}$ and computed the capture rate for different stellar velocities. The results are shown in Fig. \ref{fig-velocities}(c). As expected, the stars that encounter less energetic WIMPs (those travelling at small velocities) capture the DM particles more efficiently.

Other velocity distributions of the DM particles may be also considered. As a matter of fact, the M-B distribution is not an accurate description of the velocity distribution in the Milky Way, as it corresponds to an isotropic isothermal sphere with a DM density profile $\rho_{\chi}\propto r^{-2}$, while both observations and simulations indicate other more plausible density profiles \citep{art-Navarroetal2010MNRAS,art-Merrittetal2006AJ}. Better fits to the data are deviations from the Gaussian distribution (some examples can be found in Refs. \cite{art-SerpicoBertone2010PhRevD,art-Scottetal2009MNRAS}) or the Tsallis distribution \citep{art-Hansenetal2006JCAP}. Departures of the Maxwellian velocity distribution have been extensively studied to derive uncertainties for direct detection experiments \citep{art-FairbairnSchwetz2009JCAP,art-McCabe2010PhRevD,art-Green2010JCAP}, and will not be repeated here. These works found that more realistic descriptions for $f(v)$ may lead to deviations of $\sim10\%$ in the signal expected on the detectors. 

\section{Stellar physics and\\the capture rate
\label{sec-stellphys}}

\subsection{$\mathbf{\sigma_{\chi,SD}\gg\sigma_{\chi,SI}}$ case}
Throughout this section we assume as our fiducial values the maximum WIMP-nucleon scattering cross sections allowed by limits from direct detection experiments. In this scenario, the SD collisions of the DM particles with hydrogen are responsible for almost all the captured DM particles. In fact, the next element in importance for a star of 1 M$_{\odot}$ in the MS is oxygen, which is more than $10^4$ times less efficient capturing DM particles than hydrogen.

\subsubsection{Capture rate over stellar life}
\begin{figure}[!t]
\centering
 \includegraphics[]{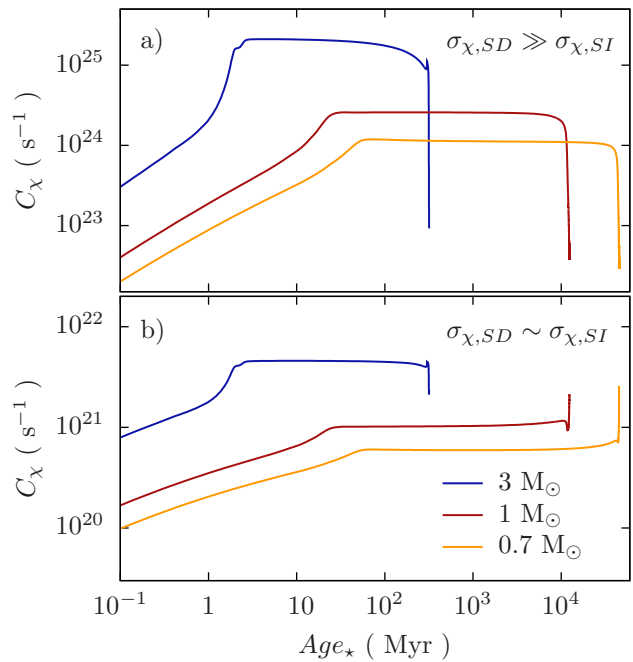}
 \caption{Rate at which DM particles are captured during the life of stars with different masses. The capture rate increases during the pre-MS, is constant through the MS, and varies rapidly in the RGB. We assumed a halo of DM particles with $\rho_{\chi}=0.3\;$GeV$\;$cm$^{-3}$, $m_{\chi}=100\;$GeV,  and the DM-nucleon scattering dominated (a) by the SD component $\sigma_{\chi,SD}=10^{-38}\;$cm$^2$ and (b) by the SI one $\sigma_{\chi,SI}=\sigma_{\chi,SD}=10^{-44}\;$cm$^2$.}
\label{fig-life}
 \end{figure}
\begin{figure}[!t]
\centering
 \includegraphics[]{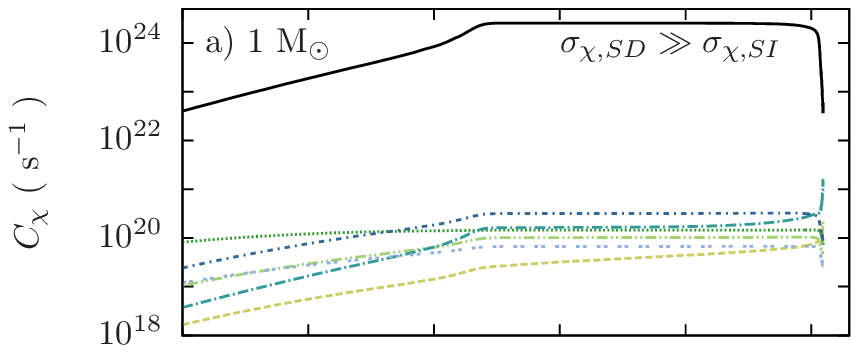}
 \includegraphics[]{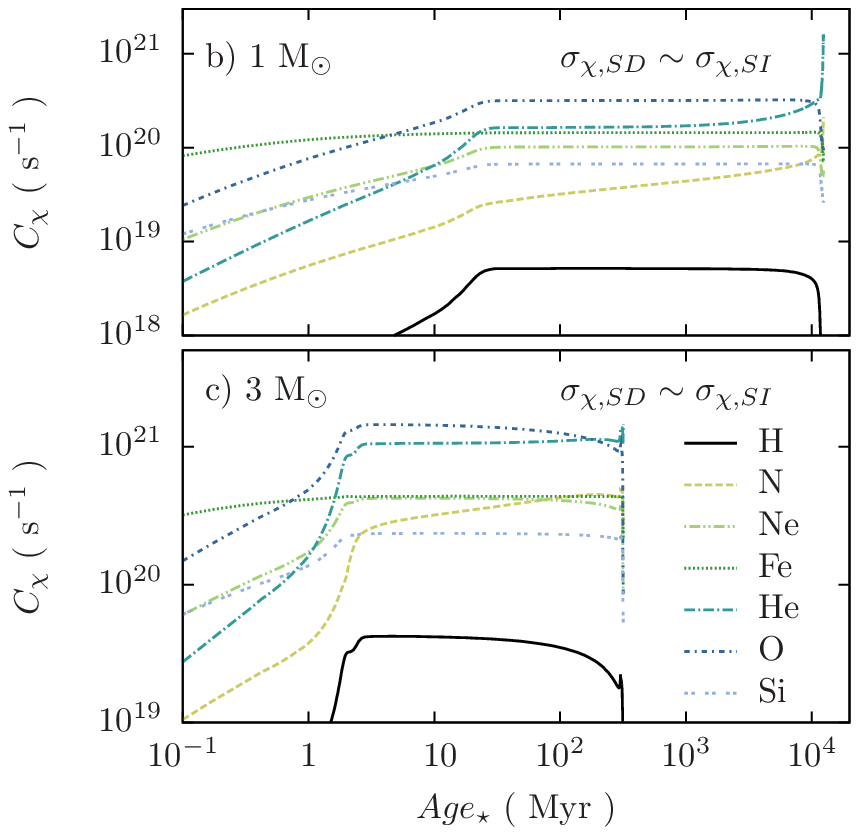}
 \caption{Rate at which DM particles are captured, discriminated by the elements responsible for the collisions that led to the capture, during the life of stars with 1 M$_{\odot}$ and 3 M$_{\odot}$. We assumed a halo of DM particles with $\rho_{\chi}=0.3\;$GeV$\;$cm$^{-3}$, $m_{\chi}=100\;$GeV, and the DM-nucleon scattering dominated (a) by the spin-dependent (SD) component, $\sigma_{\chi,SD}=10^{-38}\;$cm$^2$ and (b and c) by the spin-independent (SI) component, $\sigma_{\chi,SI}=\sigma_{\chi,SD}=10^{-44}\;$cm$^2$.}
\label{fig-elemSD}
 \end{figure}
The evolution of the capture rate through the life of the star is studied in this section. Normally, a constant capture rate is assumed during the MS, and it is expected to vary rapidly during the pre and post MS phases due to the changes in the stellar structure. To address this question in detail, the capture rate was also computed during the gravitational collapse of the protostar and during the red giant branch (RGB) until the helium flash. The results are shown in Fig. \ref{fig-life}(a) for stars with different masses.

As expected, we found that the capture rate increases continuously as the protostar collapses, remains constant during the MS, and finally drops suddenly in the RGB, when the star expands with hydrogen fusion undergoing only in a shell out of the contracting helium core. The changes in the capture rate mimic the changes in the global properties of the star, in particular in the radius of the star and in the density of the various stellar elements  $n_i(r)$, specially hydrogen.

However, as shown in Fig. \ref{fig-elemSD}(a), the predominance of hydrogen is reduced to just an order of magnitude in the RGB. At this stage the $^4$He, produced in the center of the star through the proton-proton chain during the MS, now forms an inert helium core with a density that increases dramatically as the star evolves through the RGB. Therefore, the efficiency of this isotope in capturing DM particles increases, getting closer to hydrogen, much more abundant in the rest of the star and still responsible for most of the captures. Another isotope that gains importance during the RGB is $^{14}$N, which is produced during the CNO cycle.

\subsubsection{Capture and stellar metallicity}
\begin{figure}[!t]
\centering
 \includegraphics[]{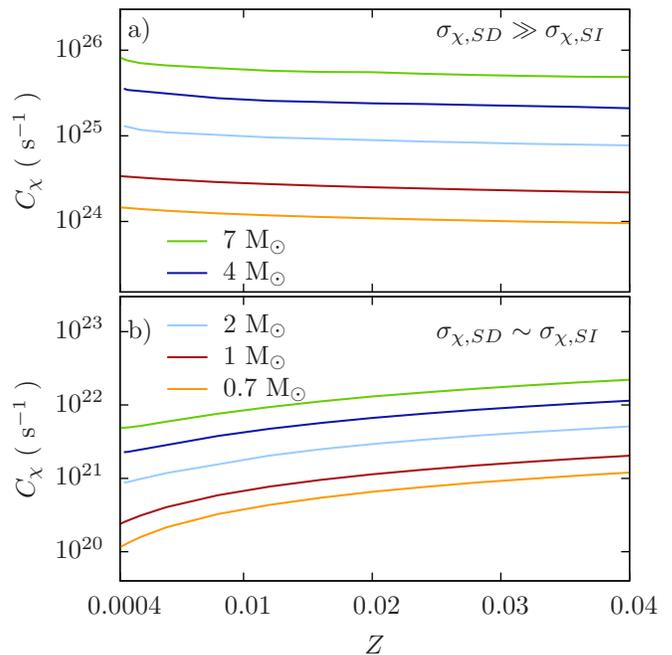}
 \caption{Rate at which DM particles are captured during the MS for stars with different masses and metallicities. We assumed a halo of DM particles with $\rho_{\chi}=0.3\;$GeV$\;$cm$^{-3}$, $m_{\chi}=100\;$GeV,  and the DM-nucleon scattering dominated (a) by the SD component, $\sigma_{\chi,SD}=10^{-38}\;$cm$^2$, and (b) by the SI one, $\sigma_{\chi,SI}=\sigma_{\chi,SD}=10^{-44}\;$cm$^2$.}
\label{fig-elemSDSI-Z}
 \end{figure}

Stars with metallicities from Z=0.0004 to Z=0.04 (with their corresponding helium mass fractions from Y=0.2412 to Y=0.340, along the lines of Refs. \cite{art-LejeuneShaerer2001,art-Schaereretal1993}) were considered in order to study the dependence of the capture rate on the stellar metallicity. As expected, stars with a reduced hydrogen mass fraction (those richer in metals), capture DM particles less efficiently [see Fig. \ref{fig-elemSDSI-Z}(a)]. However, regarding the importance of DM annihilation inside stars this drop on the capture rate is too small and is balanced by the fact that metal-rich stars also produce energy through thermonuclear reactions at a lower rate \citep{art-CasanellasLopes2009ApJ}.

\subsection{$\mathbf{\sigma_{\chi,SD}\sim\sigma_{\chi,SI}}$ case}
\begin{figure}[!t]
\centering
 \includegraphics[]{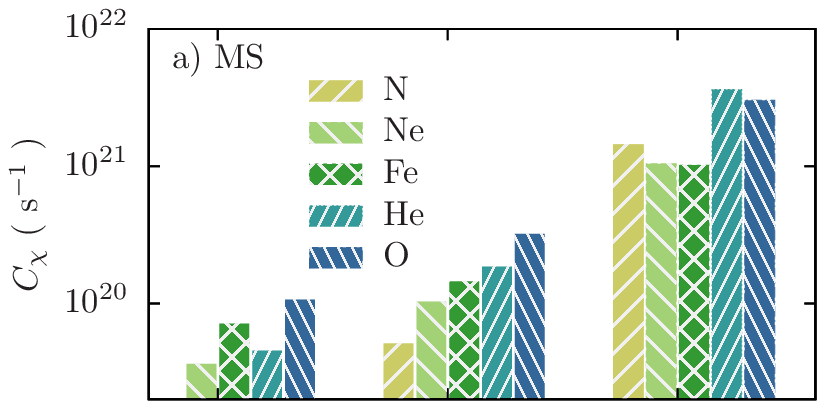}
 \includegraphics[]{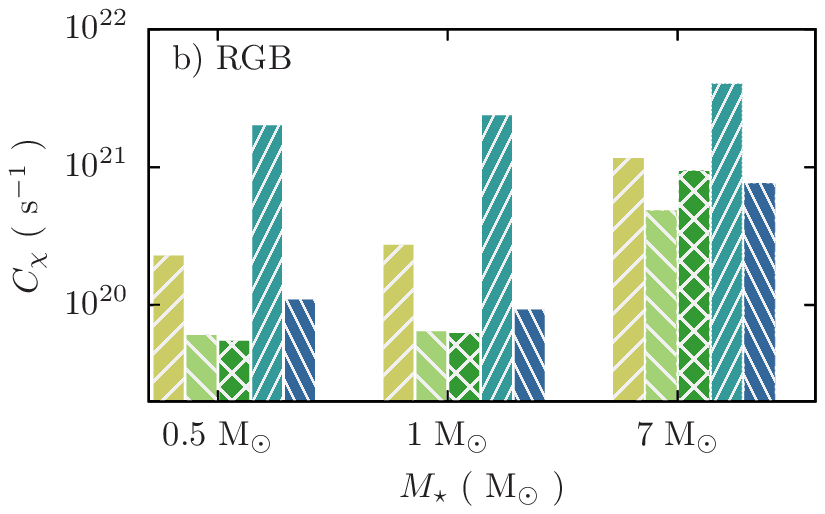}
 \caption{Rate at which DM particles are captured discriminated by the elements responsible for the collisions that led to the capture, in the main sequence (a), and in the red giant branch (b) for stars with different masses. We assumed a halo of DM particles with $\rho_{\chi}=0.3\;$GeV$\;$cm$^{-3}$, $m_{\chi}=100\;$GeV,  and the DM-nucleon scattering dominated by the spin-independent (SI) component, $\sigma_{\chi,SI}=\sigma_{\chi,SD}=10^{-44}\;$cm$^2$.}
\label{fig-elemSI-mass}
 \end{figure}
\begin{figure}[!t]
\centering
 \includegraphics[]{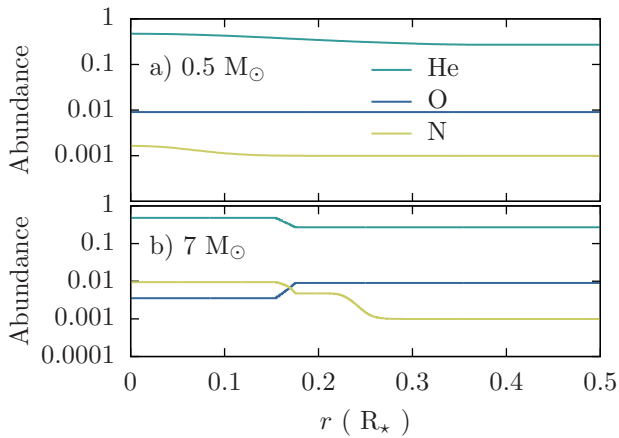}
 \caption{Radial abundances of $^4$He, $^{16}$O and $^{14}$N for stars of 0.5 M$_{\odot}$(a) and 7 M$_{\odot}$ (b) in the middle of the MS (when $X_c=0.5$).}
\label{fig-abundSI}
 \end{figure}

A scenario in which the SD and SI scattering cross sections have similar values is also considered in this section. In fact, normally a larger SD cross section is assumed because the limits from detectors are less stringent, due to technological limitations. But, as the processes leading to these interactions are similar, both scattering cross sections are of the same order in most models if no resonances nor destructive interferences are invoked \citep{rev-JungmanKamionkowskiGriest1996}. Thus, we choose $\sigma_{\chi,SD}=10^{-44}\;$cm$^2$ and $\sigma_{\chi,SI}=10^{-44}\;$cm$^2$ in order to explore in depth the role of the different stellar elements in the capture of DM particles.

In this case, and also even if we had chosen a $\sigma_{\chi,SD}$ up to 2 orders of magnitude greater than $\sigma_{\chi,SI}$ (for a star with $Z \sim Z_{\odot}$), the SI interactions are the dominant ones in capturing DM particles. The most important elements for the total capture rate, in a star of 1 M$_{\odot}$ during the MS, are oxygen, helium, iron, and neon. The heavier elements, such as iron, do not dominate the capture rate owing to the form-factor suppression.

Stars of different masses may have other elements contributing significantly to capturing DM particles. For instance, in a star of 7 M$_{\odot}$ in the MS, helium is the most important element, followed by oxygen and nitrogen [see Fig. \ref{fig-elemSI-mass}(a)]. On the other hand, in a star of 0.5 M$_{\odot}$ oxygen arises as the element that captures DM particles more efficiently, followed by iron. These different contributions are explained by the abundances of the elements throughout the star [see Fig. \ref{fig-abundSI}]. Some of the $^{16}$O in a star of 7 M$_{\odot}$ is converted to $^{14}$N through the CNO cycle, while the same does not happen for a star of 0.5 M$_{\odot}$ \citep{book-Bohm-Vitense1992}.

\subsubsection{Capture rate over stellar life}
The importance of helium and nitrogen on the capture rate increases at the final stages of evolution, in opposition to the cases of hydrogen and iron, whose contribution drops in the RGB [see Figs. \ref{fig-elemSD} and \ref{fig-elemSI-mass}]. As a consequence, when the SI interactions dominate, the capture rate does not drop so abruptly in the RGB. Moreover, we found that for stars with masses smaller than 2 M$_\odot$ the total capture rate increases in the RGB instead of decreasing [see Fig. \ref{fig-life}(b)]. Although the number of captured DM particles increases in the RGB, the influence of their self-annihilation on the stellar properties is not remarkable, as at the same time the energy from thermonuclear reactions also increases dramatically.

When the stars are in the RGB the elements responsible for most of the DM captures are different from those on the MS [Fig. \ref{fig-elemSI-mass}]. In the RGB, helium is the most important element for all stars with masses in the range 0.5 M$_{\odot}$-7 M$_{\odot}$. The huge density reached by the helium core in the RGB ($\rho_{c,RGB}\sim10^3\rho_{c,MS}$) increases the efficiency of this element in capturing DM particles.

It is also remarkable that, in the pre-MS phase, the capture rate is not so small when compared with the one in the MS. In the scenario where the capture rate due to SD scattering dominates $C_{\chi,PMS}\sim 1/20\;C_{\chi,MS}$ while, if both scattering cross sections are of the same order, then $C_{\chi,PMS}\sim 1/4\;C_{\chi,MS}$ [see Fig. \ref{fig-life}]. The explanation of this fact is found in the role of iron, which is the more efficient element in capturing DM particles in the pre-MS phase [see Fig. \ref{fig-elemSD}(b) and (c)]. For most of the stellar isotopes the capture process is ineffective due to the small escape velocity inside the protostar. However, the kinetic factor in the capture rate expression is not so strongly suppressed for those isotopes with heavy nuclear masses, and therefore the elements with a large $A_i$, as iron, are the more efficient ones capturing DM particles in the pre-MS phase.

\subsubsection{Capture and stellar metallicity}
\begin{figure}[]
\centering
  \includegraphics[]{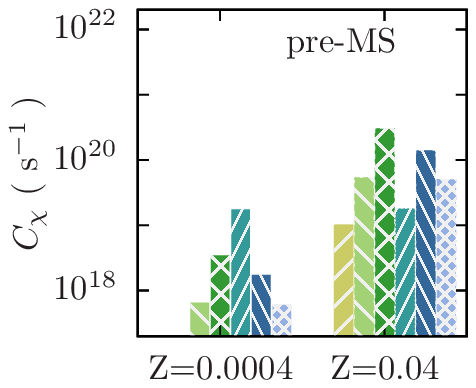}
  \includegraphics[]{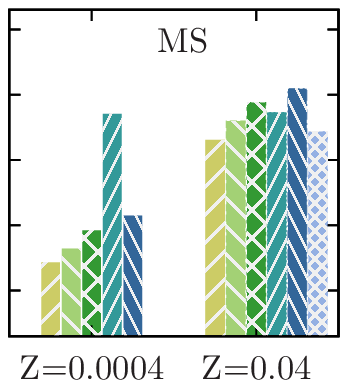}
  \includegraphics[]{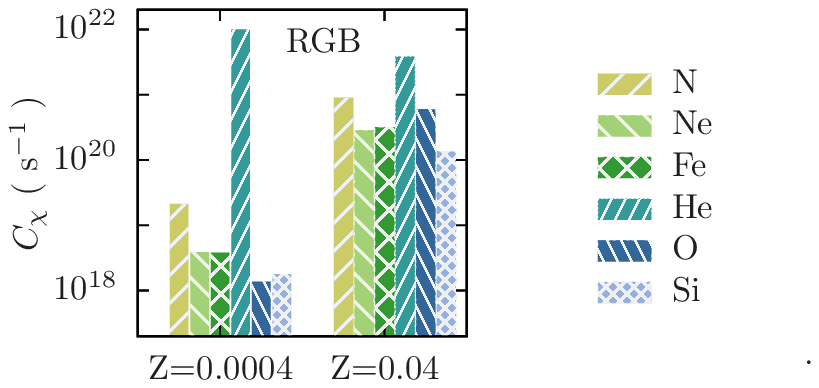}
 \caption{Rate at which DM particles are captured discriminated by the elements responsible for the collisions that led to the capture, for 1 M$_{\odot}$ stars in different stages of evolution (pre-MS, MS and RGB) and considering different stellar metallicities. We assumed a halo of DM particles with $\rho_{\chi}=0.3\;$GeV$\;$cm$^{-3}$, $m_{\chi}=50\;$GeV and the DM-nucleon scattering dominated by the spin-independent (SI) component, $\sigma_{\chi,SI}=\sigma_{\chi,SD}=10^{-44}\;$cm$^2$.}
\label{fig-elemSI-Z}
 \end{figure}
In contrast to what is expected when the SD interactions dominate, in this scenario we found that stars with higher metallicities capture DM particles more efficiently [see Fig. \ref{fig-elemSDSI-Z}(b)], because these stars are richer in the isotopes that are responsible for most of the captures: $^{16}$O, $^4$He, $^{56}$Fe, $^{20}$Ne and $^{14}$N. Therefore, in this scenario, stars with higher metallicities are more affected by the capture and annihilation of DM particles in their interior. Moreover, as metal-rich stars have lower thermonuclear energy production rates, the energy from DM annihilation is even more important over the total energy of the star (the ratio $L_{\chi}/L_{nuc}$ for a 1 M$_{\odot}$ star with Z=0.04 in the MS is more than 20 times greater than that for the same star with Z=0.0004).  

The contribution of the metals in the capture rate is of vital importance for stars with metallicity Z=0.04, especially until the end of the MS, while for stars with Z=0.0004 helium dominates the capture during all the stages [see Fig. \ref{fig-elemSI-Z}]. On the other hand, on the RGB the role of the metals is less important because in this phase $^4$He is the isotope that captures DM particles more efficiently due to its high density in the core.

\section{Discussion}
\begin{table}
 \caption{Variations in the total capture rate, $C_{\chi}$, and in the ratio between the luminosities from DM annihilations and thermonuclear reactions, $L_{\chi}/L_{nuc}$, when there is an uncertainty of 10\% in the knowledge of one parameter of the DM characteristics or of the stellar structure. If not stated otherwise, we assumed a halo of DM particles with a mass $m_{\chi} = 100$ GeV, a velocity dispersion $\bar{v_{\chi}} = 270$ km s$^{-1}$, and a star of 1 M$_{\odot}$ in the middle of the MS, with a metallicity Z=0.019 and a velocity $v_{\star} = 220$ km s$^{-1}$.}
\label{tab-uncertCx}
 \begin{ruledtabular}
 \begin{tabular}{l |r r| r r}
 & \multicolumn{2}{l|}{$C_{\chi}$} & \multicolumn{2}{l}{$L_{\chi}/L_{nuc}$}\\
\hline
$m_{\chi} = 5$ GeV $\pm 10\%$ & $-10\%$ & $+12\%$ & $-1\%$ & $+1\%$ \\
$m_{\chi} = 500$ GeV $\pm 10\%$  & $-18\%$ & $+23\%$ & $-9\%$ & $+11\%$ \\
\hline
$\bar{v_{\chi}} = 100$ km s$^{-1}$ $\pm 10\%$ & $+6\%$ & $-7\%$ & $+6\%$ & $-7\%$ \\
$\bar{v_{\chi}} = 500$ km s$^{-1}$ $\pm 10\%$ & $-20\%$ & $+26\%$ & $-20\%$ & $+26\%$ \\
\hline
$v_{\star} = 100$ km s$^{-1}$ $\pm 10\%$ & $-3\%$ & $+3\%$ & $-3\%$ & $+3\%$ \\
$v_{\star} = 500$ km s$^{-1}$ $\pm 10\%$ & $-58\%$ & $+120\%$ & $-58\%$ & $+120\%$ \\
\hline
$M_{\star} = 0.5$ M$_{\odot}$ $\pm 10\%$ & $+26\%$ & $-22\%$ & $-20\%$ & $+26\%$ \\
$M_{\star} = 7$ M$_{\odot}$ $\pm 10\%$ & $+16\%$ & $-13\%$ & $-16\%$ & $+26\%$ \\
\hline
$Z = 0.0004 \pm 10\%$ & $-0.1\%$ & $+0.1\%$ & $+2\%$ & $-0.3\%$ \\
$Z = 0.04 \pm 10\%$ & $-2\%$ & $+2\%$ & $-2\%$ & $+1\%$ \\
 \end{tabular}
\end{ruledtabular}
\end{table}
\begin{table}
 \caption{Variations in the capture rate due to SD and SI interactions of the stellar elements with the DM particles ($C_{\chi,SI}$ and $C_{\chi,SD}$) when there is an uncertainty of 10\% in the knowledge of the mass of the DM particles or on the stellar metallicity.}
\label{tab-uncertCxSDSI}
 \begin{ruledtabular}
 \begin{tabular}{l | l l | l l}
  & \multicolumn{2}{l|}{$C_{\chi,SD}$} & \multicolumn{2}{l}{$C_{\chi,SI}$}\\
\hline
$m_{\chi} = 100$ GeV $\pm 10\%$ & $-16\%$ & $+22\%$ & $-10\%$ & $+13\%$ \\
\hline
$Z = 0.019 \pm 10\%$ & $-2\%$ & $+2\%$ & $+8\%$ & $-8\%$ \\
 \end{tabular}
\end{ruledtabular}
 \end{table}

We have characterized how the stellar capture of DM particles changes within different assumptions regarding the DM characteristics and the structure of the stars. These results are summarized in Table \ref{tab-uncertCx}, where we show the variations in the computed capture rate derived from an uncertainty of 10\% in the knowledge of given parameters (such as the mass and velocities of the star and the DM particles, and the stellar metallicity). We found that the greater uncertainties in the capture rate occur due to the ignorance of the DM particle mass and specially when the stellar velocity (if very high) and the stellar mass are not well determined.

However, not all uncertainties in the computed capture rate contribute equally to the weight of the subsequent DM annihilations over the nuclear sources of energy of the star. To illustrate this fact the ratio $L_{\chi}/L_{nuc}$ is also shown in the third column of Table \ref{tab-uncertCx}. In this respect, the ignorance of the DM mass is much less important when compared with an imprecise determination of the velocities or the stellar mass. As an example, an overestimation of 10\% in the mass of a star of 7 M$_{\odot}$ leads to a significant increment on the computed capture rate (+16\%), while regarding the effects of DM annihilation on the same star, this overestimation is completely counterbalanced by the dependence of the thermonuclear energy sources on the stellar mass.

The errors on the estimation of the stellar metallicity are not significant for the computed capture rate, at least for the SD-dominated capture. In the scenario where the SI interactions dominate, the role of the metallicity is more important but still introduces errors on the capture rate below 10 \% (see Table \ref{tab-uncertCxSDSI}).

The relatively large variations on the computed capture rate due to a poor knowledge of the input physics stress the importance of combining different techniques to improve precision in the determination of the parameters. In the case of the stellar parameters, photometry, spectroscopy and astroseismology should be combined when possible to reduce the uncertainties in the stellar mass and metallicity. Regarding the DM characteristics, only a combination of results from colliders, direct and indirect detection experiments will constrain sufficiently the free parameter space. In the cases where the detection of DM signatures seems more promising, such as the Galactic center and primordial stars, the uncertainty on the capture rate will be dominated by the ignorance on the exact value of the DM density.

Our results may be used to evaluate the reliability of the computed capture rate for stars observed in environments with high expected DM densities, and therefore to estimate if the effects predicted due to the self-annihilation of DM particles in the stellar interiors will allow us to extract information about the nature of DM.

\begin{acknowledgments}
We acknowledge the authors of {\sf CESAM} (P. Morel) and {\sf DarkSUSY} (P. Gondolo, J. Edsj\"{o}, P. Ullio, L. Bergstr\"{o}m, M. Schelke and E. Baltz). Furthermore, we thank J.Edsj{\"{o}} for fruitful comments on the subject of the capture of dark matter particles in the Sun. This work was supported by grants from ``Funda\c c\~ao para a Ci\^encia e Tecnologia" (SFRH/BD/44321/2008) and "Funda\c c\~ao Calouste Gulbenkian".
\end{acknowledgments}

\bibliography{DM}

\end{document}